# A Compendium on Distributed Systems

Aneesh Khole
*Computer Science Dept*
*(of MMCOE)*
Marathwada Mitra Mandal's College of Engineering
*(of SPPU)*
Pune, India
aneeshkhole2019.comp@mmcoe.edu.in

Atharva Thakar
*Computer Science Dept*
*(of MMCOE)*
Marathwada Mitra Mandal's College of Engineering
*(of SPPU)*
Pune, India
atharvathakar2019.comp@mmcoe.edu.in

Avadhoot Kulkarni
*Information Technology Dept*
*(of PVGCOET)*
Pune Vidhyarthi Griha's College of Engineering and Technology
*(of SPPU)*
Pune, India
avadhootkulkarni06@gmail.com

Hrithik Jadhav
*Mechanical Dept*
*(of MMCOE)*
Marathwada Mitra Mandal's College of Engineering
*(of SPPU)*
Pune, India
hrithikjadhav2019.mech@mmcoe.edu.in

Shreyas Shende
*Computer Science Dept*
*(of MMCOE)*
Marathwada Mitra Mandal's College of Engineering
*(of SPPU)*
Pune, India
shreyasshende2019.comp@mmcoe.edu.in

Varad Karajkhede
*Computer Science Dept*
*(of MMCOE)*
Marathwada Mitra Mandal's College of Engineering
*(of SPPU)*
Pune, India
varadkarajkhede2019.comp@mmcoe.edu.in

*Abstract*— Computer systems have evolved over the years starting from sizable, single-user, slow, and expensive machines to multi-user, fast, cheaper, and small-sized machines. The use of multi-user computer networks has given rise to a new paradigm of computing known as Distributed Systems. A distributed system is regarded as software consisting of a collection of dependent network communication and computational nodes. This paradigm yields high performance while also maintaining high efficiency due to the decentralization of various computer-related tasks to several computer nodes that are interconnected. Even if distributed systems have proven to be beneficial over the years it also has some design flaws, security concerns and challenges. In this paper, the main objective is to define these issues, challenges and security concerns while also examining the various solutions developed over the years to resolve them. This paper also briefly covers the components as well as the working of Distributed Systems.

## I. INTRODUCTION

1.1 What is a Distributed System?

Several definitions of Distributed Systems have been presented over the years, but neither are they satisfactory nor in agreement with one another. Among some definitions by Coulouris, there is one that defines a distributed system as *"a system in which hardware or software components located at networked computers communicate and coordinate their actions only by message passing".* [1]

Another definition to give a brief overview of Distributed Systems is:

A distributed system is a computing environment in which numerous components are located across multiple computing devices on a network. These devices segregate the work, communicating and coordinating their efforts to appear to be a single cohesive system to the end-user. A distributed system can be an arrangement of different configurations, such as mainframes, computers, workstations, and minicomputers.

This definition highlights two principal features of distributed systems. The first one is the idea that a distributed system is made up of various computer components that can include either software or hardware components. They are autonomous of each other, where the computing elements are referred to as nodes. The second characteristic is that users feel like they are operating with the system alone.

1.2 Why do we need Distributed Systems?

Modern computing is possible because of distributed systems. They are essential for the performance of cloud computing services, wireless networks and the Internet. Distributed systems offer better availability. Without distributed systems, these advances would not have been possible. In most cases, distributed systems are even used in corporate positions that lack the complexity of an entire telecommunications network. Distributed systems overcome the drawbacks of monolithic systems; they provide more scalability and improved performance.

Distributed systems can provide features that can be difficult to upgrade for a single system because they can use other computer devices and processes. Performing regular backups of servers and applications, including those located at remote locations, is an essential task. In the event of a restore, the necessary data can be obtained from the off-site backup nodes by requesting the relevant segments.

Everything on a computing device works on the power of distributed systems virtually.

## II. ARCHITECTURE AND WORKING OF DISTRIBUTED SYSTEMS

2.1 Elements of Distributed Systems

The three components of a distributed system include a primary system controller, system data store, and database. When it comes to non-clustered environments optional components such as user interfaces and secondary controllers can be used.

2.1.1 Primary System Controller

The primary system controller is the only present controller in the distributed system that keeps everything. It is accountable for management as well as controlling the dispatch of server requests throughout the system. The primary system controller keeps track of status information for the comprehensive system and has control over all other system components. The installation of executive and mailbox services automatically takes place on the primary system controller.

2.1.2 Secondary Controller

The secondary controller acts as a process or a communication controller. It's responsible for regulating the flow of server processing requests and managing the system's translation load. It is also responsible for governing the communication between systems and VANs or trading partners.

2.1.3 User-Interface client

The user interface client is also an additional element in the system whose task is to provide important system information to the users. The user interface controller isn't a part of a clustered environment and doesn't operate on the same machines as the controller. Its functions include monitoring and controlling the system.

2.1.4 System datastore

The system's data store is normally present on the disk vault whether clustered or not and each system has only one data store for all shared data. In non-clustered systems, the data store can be present on a single machine or across several machines, but all the computers should have access to this datastore.

2.1.5 Database

For a distributed system all the data gets stored in a relational database. After locating the data from the database, the data store shares the data among multiple users. All the data systems have relational databases and it allows multiple users to use the same information at the same time.

2.2 Architectures of Distributed Systems

The `system-level architecture` focuses on the entire system and the placement of components of a distributed system across multiple machines. Distributed systems are divided into four different architecture models

1. Client-server - Client-server architecture is composed of a client and a server. Data is retrieved from the server by clients, who subsequently format it and display it to the user. The end-user can also make amends from the client-side and restore them to the server to make them permanent. Client-server architecture has one standard design feature: centralized security.

2. Three-tier - The clients in a three-tier architecture are stateless. To simplify the application deployment, the information rather than being kept on the client side, the client information is kept in a middle-tier. Three-tier architecture is most common for web applications.

3. *N*-tier - In n-tier architecture, the middle layer interacts with another service to get information. N-tier architecture is usually used when an application or server needs to forward requests to additional enterprise services on the network.

4. Peer-to-peer - A peer-to-peer network, also called a (P2P) network, works on the concept of a decentralized system. There are no additional machines used to provide services or manage resources. The machines in the system, called peers, have uniformly distributed responsibilities. Both clients and servers can use them.

2.3 Working of Distributed Systems

Distributed systems have evolved with today's use cases of distributed systems which are designed to operate over the internet and cloud. The working of distributed systems can be explained through this example. The process begins with a task like rendering a video. The video editor then manages this task and splits the work into pieces. Multiple computers (referred to as nodes) are used with each node getting a frame of the video. These nodes work independently on a single frame and when they complete the rendering on that frame a new frame is given by the video editor to work on. This process goes on until the rendering of all frames is completed. Also due to the distributed system, there is no limitation on how many nodes can be used. The quantity of nodes we use is directly proportional to the time required to complete the rendering process. Therefore, using more nodes means the process is finished faster. It is described in the diagram below.

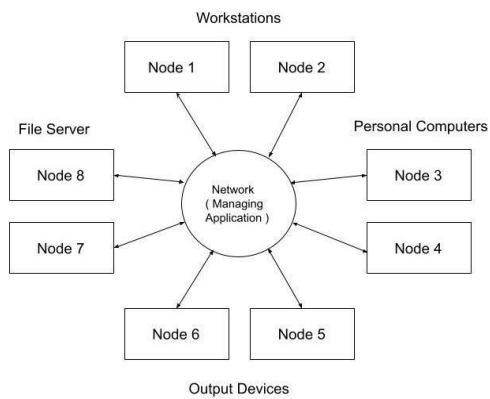

Figure 1. Illustration Of Distributed System

All the nodes are connected through the network and they communicate with each other while the process is running. These nodes work in tandem to complete the process while also being able to perform the operations. Several models and architectures are being used today like client-server and peer-to-peer systems. Both of these architectures produce the same output but their methods to achieve the output are different. The working of a distributed system is explained by the above example.

## III. SECURITY

Security in distributed systems can kind of be divided into two parts. One element guides the communication among users or processes, probably residing on special machines. The important mechanism for ensuring secure communication is that of a secure channel. The second part concerns authorization, which offers with ensuring that a process receives the access rights to only the resources of the distributed system. It is concerned with secure channels and access control that requires mechanisms to distribute cryptographic keys, but also mechanisms to add and put off users from a system. These subjects are covered using what is called security management.

Protection in a computer machine is firmly related to the idea of dependability. Unofficially, a dependable computer system is one which we believe to provide its services. Dependability in a computer system encompasses various elements such as its accessibility, stability, safety, and ease of maintenance. However, to ensure that the system can be fully trusted, it is also crucial to consider factors such as the protection of sensitive information and the preservation of data integrity. Wrong alterations in a secure computer system must be detectable and recoverable. The most important assets of any computer system are its hardware, software program, and statistics.

3.1 Security Threats

1. Interception

The concept of interception points out the circumstance that an unauthorized entity has acquired access to a service or some information.

2. Interruption

Interruption is the condition in which services or data become hard to get, unusable, corrupt, etc. According to the `denial of service attacks`, by which some third party illegally attempts to make a service inaccessible to other entities is a security threat that identifies as an interruption.

3. Modification

Modifications imply the unauthorized sabotaging of data or tampering with a service so that it no longer coheres to its formal specifications.

4. Fabrication

Fabrication is the condition in which extra data or activity are created that would normally be absent.

3.2 Security Requirements

1. Confidentiality

It consists of data being inaccessible to unauthorized individuals. Only after the authentication process, data is kept accessible to concerned authorities.
Confidentiality is maintained by encryption of data.

2. Integrity

It avoids any unauthorized changes to the data and detects if any changes are made. Many authentication algorithms are used for such validation processes. This helps only authorities to modify any piece of data.

3. Availability

It provides the availability of data only to the concerned authorities. This means that data is not available to any person to modify or change and hence is a security requirement.

3.3 Security Mechanisms

1. Encryption

Encryption is fundamental to computer security. Encryption transforms data into an entity the attacker cannot interpret easily. Encryption provides a means to apply data confidentiality and allows us to check whether data have been modified. The primitive approaches are:

i. Conventional Encryption: Here the same key is used by a sender to encrypt a message and by the receiver to decrypt the message. It is also known as Symmetric Encryption.
ii. Public-key Encryption: Here two keys are used for encryption purposes and one key is made public for anyone to use. It is also known as Asymmetric Encryption.

2. Authentication

Authentication is used for verification of the claimed identity of a user, client, server, host, or other entity. For clients, the basic thing is that before the service starts to perform any work on the pretext of a client, the service must learn the client's identity.

3. Authorization

After a client has been authenticated, it is necessary to check whether that client is authorized to perform the action requested. It means checking if the client has the requirements satisfied.

4. Auditing

Auditing tools are used to track which clients accessed what, and in which way. Even if this method does not provide any protection against threats, audit logs can be exceedingly important for the study of a security violation, and accordingly taking measures against attackers.

3.4 Use of Cryptography in Security Measures

The use of cryptographic techniques is an important fundamental to security in distributed systems. Its use cannot ensure the security of a system but it is a very important component for building secure distributed systems.
The application of these techniques is very simple.
Consider a situation where a sender wants to transmit a message to another receiver. The sender converts the same message into an unintelligible message and sends it to the recipient in order to protect the message from various security concerns. For obtaining the message in its original form, the receiver must decrypt the message. Such encryption-decryption methods are made possible by cryptographic methods parameterized by keys. Here the original message is termed plaintext whereas its encrypted form is termed ciphertext.
There are three different attacks that we need to protect against while transferring the ciphertext message. The first possible threat is an intruder, intercepting the message without the sender or receiver being aware. If the plaintext is encrypted in such a way that decryption is not possible with the appropriate key, interception is of no point. The intruder will be able to view only unintelligible data. The second type of possible attack is modifying or making any changes to the message. Although modifying the plain text is not difficult, making changes to ciphertext which has been encrypted is tougher because the message needs to be decrypted first to modify it meaningfully. Also, the encryption post modifying the message has to be accurate, otherwise, the recipient might notice that the message has been meddled by some unauthorized personnel.
The last type of attack which is possible is when the intruder inserts encrypted messages by itself and the receiver thinks that the messages are being sent by the sender. [2]

3.5 Design Issues

A distributed system, or any computer system, has to provide security systems by which a huge variety of safety guidelines may be carried out. When imposing general-purpose security offers, there are a few crucial design considerations that must be made. While considering the protection of an application (possibly distributed), three unique methods may be observed.

1. The first approach is to concentrate on the protection of the data and information that is associated directly with the application. By direct, it means that irrespective of the various functions that can potentially be performed on a data item, the primary concern is to ensure data integrity. Generally, this form of security happens in database systems in which numerous integrity constraints can be formulated that are routinely checked on every occasion a data item is altered.

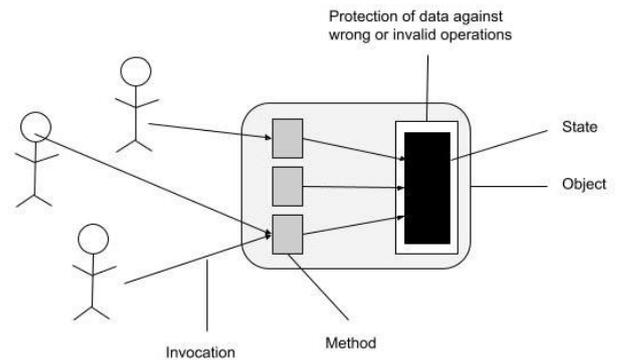

Figure 2. Protection Of Data Against Invalid Operations *[2]*

2. The second approach is to concentrate on security by specifying exactly which operations may be called, by whom, and when certain data or resources are to be accessed. In this case, the power of control is strongly related to access control methods. Access control methods can be applied to an entire interface offered by an entity, or to the entire object itself. This approach, therefore, allows for various granularities of access control.

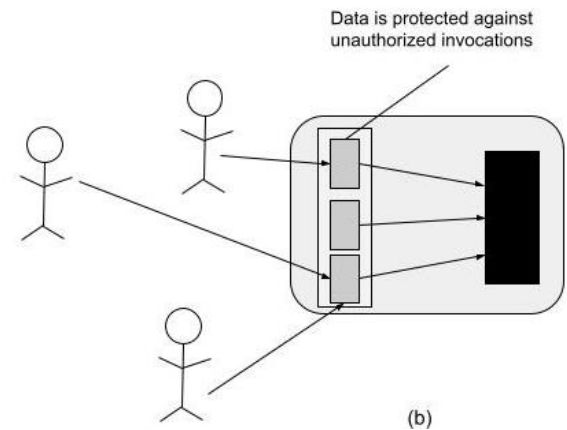

Figure 3. Protection Of Data Against Unauthorized Invocations *[2]*

3. A third approach is to concentrate directly on users by taking measures by which only selected personnel to have access to the application, irrespective of the functions they want to carry out. In effect, control is focused on defining use cases that users have, and once a user's role has been verified, access to a resource is either granted or denied. As part of designing a secure system, it is thus necessary to characterize roles that people may receive, and provide mechanisms to support role-based access control.

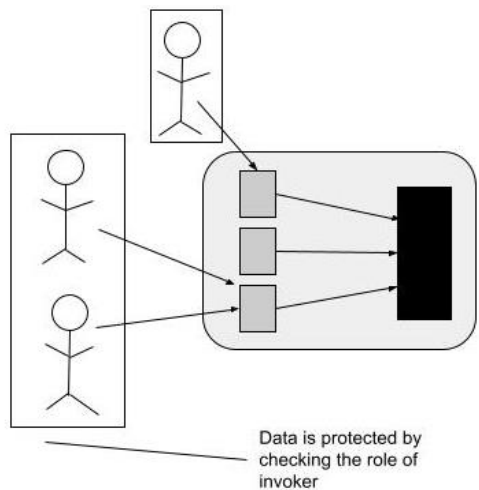

Figure 4. Protection Of Data By Checking The Role Of Invoker [2]

## IV. CHALLENGES OF DISTRIBUTED SYSTEMS

Designing a distributed system does not come as easy. Several challenges need to be overcome to get the ideal system. The major challenges in distributed systems are listed in the diagram below.

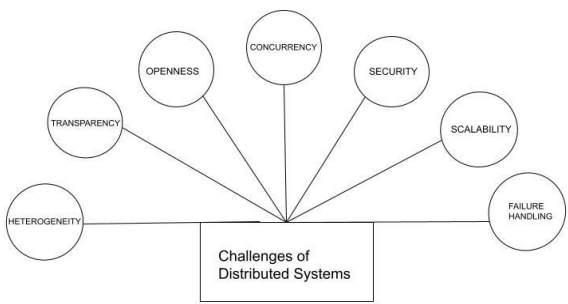

Figure 5. Challenges Of Distributed Systems

### 4.1 Heterogeneity

Providing access to various services and executing applications across a diverse set of computers. Utilities and praxis span a significantly varied group of computer networks that can be used, run and accessed by clients and end-users over the internet.
Hardware devices: computers, tablets, mobile phones, embedded devices, etc.
Different OS variants like the Windows OS, Linux OS, mac and Unix. Different types of networks that may be used include a local network, the internet, wireless networks, and satellite connections. Programming languages such as Java, C/C++, Python, and PHP can be utilized. For communication to be implemented, standardized internet protocols are required and needed to be accepted. The term middleware pertains to a tier of software that supplies a programming abstraction in addition to impersonating the heterogeneity of the core networks, programming languages, and operating systems hardware. A portable code is nothing but a program code that can be transferred from one computer system to another and can be redistributed from a single computer. One such example of a portable code is Java Web Start as the Java applet is now deprecated.

### 4.2 Transparency

The idea of a distributed system as a single unit of the centralized system rather than a congregation of independent elements is the main idea behind transparency in distributed arrangements. A distributed systems developer aims to conceal the intricacies as much as feasible. Transparency criteria may be custom permissions, removal, concurrency, break-down, duration, and customer resources. Transparency is also delineated as the isolation of components in a distributed system so that the system is considered as a whole rather than as an asset of autonomous components. In other words, distributed systems developers must conceal the complicatedness of the system as doable as they can. The motive behind Total Transparency is to completely conceal the experience that the computation is being offered by a distributed System.

### 4.3 Openness

The motive behind openness in distributed systems is to devise the network in a way that can be simply configured or modified. Developers are often required to add new features or revise and replace any existing aspect in any distributed node, which is easily facilitated if there is appropriate support for interoperability based on some standard protocols. Openness issues are exacerbated when content that has been published is either revoked or abruptly deleted. Besides, most of the time there is no central authority in distributed open systems, as different systems may have their arbitrator. For example, organizations like Meta, Tinder, etc. Allow developers to build their software interactively through their API.

### 4.4 Concurrency

Concurrency manages synchronal access to resources. It stops multiple users from making changes to the same record at the same time and also organizes transactions in a specific order for backup and recovery purposes. For example, during an auction multiple participants bid on the same object, similarly in a distributed environment, as servers may attempt to access shared resources, there is, therefore, a chance that several clients will try to access the shared service at the same time. Object to be safe in the same place, its functions must be performed in a critical environment and processes need to be harmonized in such a way that its data remains consistent. This can be achieved in ways such as the use of semaphores, which are widely used in many applications. Multiple users when trying to access the same set of data or shared resources, led to concurrency issues. This significantly changes or affects the final output of the process as without synchronization the processes get executed in the order in which they are inputted.

4.5 Security

Most of the information resources that are made availed and maintained in distributed systems have a high inherent value to their users. Their security is therefore significantly important. While public networks are being used, security is the biggest issue concerning the distributed environment. Security for information resources has three components:

1. Confidentiality- (protection against divulgence to unauthorized persons)
Integrity- (protection against alterations or deception).

2. Availability- (protection against interfering with ways to client resources), must be provided in DSS. Encryption is one of the methods to avoid security concerns. Ensuring that only authorized and legitimate users can access the resources, modify and perform operations. Presently many institutions and organizations in the world have designed and developed systems with distributed environments that possess all of the security features mentioned above.

4.6 Scalability

A system encounters scalability issues when it lacks the capabilities to handle an abrupt boost of resources and or several users. In such situations, efficient use of architecture and algorithms must be done. A system demands scaling on specifications like size, Geography, or Administration.

- Size: Size is the number of users and resources to be processed. Problems that may arise due to size include overloading.
- Geography: Geography is the distance that links users and resources. Communication reliability is one such issue that arises due to geographic limitations.
- Administration: Nodes of distributed systems need to be controlled as the dimensions steadily increase. Administrative chaos and its related difficulties arise due to scalability problems associated with administration.

4.7 Failure Handling

Computer systems may fail sometimes. When such failures occur in hardware or software, programs may produce incorrect results or may terminate before they have completed the intended task. The handling of such failures is extremely difficult. A distributed system comprises a group of autonomous computing devices that present a unified front to clients, operating as a cohesive and integrated whole. The system is composed of a variety of constituent parts that synergize in the accomplishment of a particular objective. As the system expands and scales itself vertically, As the system becomes more distributed, the chances of system failures increase, resulting in a diminution of dependability. In other words, in a distributed system, there may be some elements that are inoperable while others remain functional. This phenomenon is regarded to as partial malfunction. Due to the non-deterministic nature of time, the message and the time required to travel to its destination are also non-deterministic, therefore we say that partial failures are unpredictable. We get no acknowledgment to know whether the system has succeeded or failed. Partial failure includes node crashes or communication connectivity issues in distributed systems.

## V. SOLUTIONS TO CHALLENGES IN DISTRIBUTED SYSTEM

5.1 Heterogeneity solution

In the effort to solve issues related to heterogeneity, we show how selected approaches in the field of distributed computing handle different kinds of heterogeneity. While there are numerous systems for sharing computation, this overview presents selected frameworks that represent the full spectrum. We introduce a batch workload management system (HTCondor), a project-based volunteer and grid computing system (BOINC), a service-oriented desktop grid approach (Aneka), an enterprise grid (Entropia), and a library-based programming model for distributed computing (libWater).

The development of Distributed Systems is the main focus of this proposition. Following things of tasks are not being considered i.e. Accessibility and their nature.

We have presented frameworks with different approaches in general. Each system has a different focus point for instance - high throughput, security, or application integration.

The table below shows us how they handle different dimensions of heterogeneity.

|  | Hardware | Operating System | Programming Language | Accessibility | Nature of Tasks |
|---|---|---|---|---|---|
| HTCondor | ✗ | ✗ | ✗ | ✓ | ✓ |
| BOINC | ✓ | ✗ | ✗ | ✓ | ✗ |
| Aneka | ✗ | ✓ | ✗ | ✓ | ✓ |
| Entropia | ✗ | ✗ | ✗ | ✓ | ✓ |
| lib Water | ✓ | ✓ | ✗ | ✗ | ✗ |

*Table 1. Dimensions Of Heterogeneity [3]*

5.2 Transparency solution

5.2.1 Access Transparency

There should be no apparent difference between local and remote access methods. In other words, open communication can be kept secret. For example, from a user's point of view, access to the remote-control service as a printer should be the same as access to a local printer. From the programmer's POV, access to a remote object may be the same as access to a local object in the same category. This transparency has two parts:
Maintaining syntactical or mechanical coherence between distributed and non-distributed access, Maintaining the same semantics. Because remote semantics are very complex, especially failure methods, this means that local access has to be a minimum set. Remote access will not always look like local access because certain services may not make sense to support (for example, a complete global search for a single-factor distributed system may not make sense in terms of network traffic).

5.2.2 Location Transparency

The details of the topology of the system should not bother the user. The location of an item in the system may not be visible to the user or editor. This differs from open access in that both the design methods and the access methods can be the same. Words may not provide any space.
Users of Concurrency and Applications should be able to access shared data or items without interruption among others. This requires a much more sophisticated approach to a distributed system, as there is more realistic compatibility than a central system simulation. For example, a distributed printing service should offer the same level of access to files as a central system, avoiding unpredictable interference during printing. The replication of the system for availability or performance reasons should not impact the user, including an app editor.

5.2.3 Replication Transparency

If the system provides repetition (for availability or performance reasons) it should not affect the user. For all openness, we include an app editor as a user.

5.2.4 Fault Transparency

If software or computer components fail, this should be hidden from the user. This can be difficult to provide in a distributed system, as a failure of the lower part of the communication system is possible, and this may not be reported. As far as possible, open errors will be provided by procedures related to public access. However, when errors are found in a distributed system environment, then accessibility light may not be maintained. Methods that allow the system to hide errors may cause changes in access methods (e.g., access to trusted objects may be different from access to simple objects). It can be challenging to differentiate between a sluggish process or processor in a software program, particularly one that is network-based. This distinction is either obscured or revealed.

5.2.5 Migration Transparency

During migration of processes or data for improved performance, reliability, or to conceal differences between hosts, the user should not be aware of the changes. This is known as migration transparency.

5.2.6 Performance Transparency

Performance transparency dictates that the system's configuration should not affect the user's perception of performance. This may necessitate the utilization of advanced resource management systems. This may require sophisticated resource management systems. In cases where resources are only accessible through low-performance networks, the distinction may not be possible.

5.2.7 Scaling Transparency

Scaling transparency requires the ability for the system to expand without impacting the application algorithms. The capacity to grow and evolve is crucial for many businesses and the system should also be able to reduce in size when necessary and allocate necessary space and/or time.

5.3 Openness Solution

To make a distributed system open, it is necessary to publish a clear and well-defined interface between components. The interfaces must be standardized and new components should be easily incorporated into the existing system.

5.4 Concurrency Solution

Take it case-by-case: - The simple solution here is to add a conditional statement for this event ordering. If the next message is a link failure notification, store the notification in memory in case you become a master later.

Replicate the computation: - In a system comprising of a solitary node, there exists a unified, global sequence of events without any conflict.

Make your event handlers transactional: - Enhance your event handlers by making them transactional. Transactions enable the appearance of a group of operations as if they were executed simultaneously or not at all, providing a robust and influential capability.

Reorder events that no one will notice: -It turns out that we can achieve even better success if we use a replication model called virtual synchrony. In short, visual synchrony provides the library with three functions: join ( ) a process team, register ( ) an event host, and send ( ) an atomic broadcast message to your entire process team. Make yourself stateless: -

In a database, the `ground truth` is stored on the base. In a network, the same is stored in the routing tables of the switches themselves. It implies that the controllers' view of the network is just a soft state, i.e., we can always recover it simply by querying the switches for their current configuration.

Guarantee self-stabilization: - The previous solutions were designed to always guarantee correct behavior despite failures of the other nodes. This final solution, my personal favorite, is much more optimistic.

1. Enforce isolation among transactions.
2. Ensure database consistency by executing transactions in a manner that maintains consistency.
3. Address read-write and write-read conflicts.

5.5 Security solution

Three computer and network address requirements

- Confidentiality: Requires data to be accessible only to authorized persons.
- Integrity: Requires that only authorized teams can modify data.
- Availability: Requires data to be available from authorized groups.

5.6 Scalability solution

Following are how we can make our distributed system more scalable

- Vertical Scaling (Stronger and faster)
- Vertical Partitioning (Divide work)
- Horizontal Scaling - Application Server (Delegate and Balance)
- Caching (Respond Fast)
- Horizontal Scaling and Replication - Database Server (Create copies)

5.7 Failure Handling Solution

Fault-tolerant distributed systems often handle failures in two steps: first, detect the failure and, second, take some recovery action. A common approach to detecting failures is end-to-end timeouts, but using timeouts brings problems.